# Quarks in a polar model

V. N. Salomatov

Irkutsk State Transport University

15 Tchernyshevsky st., 664074 Irkutsk, Russia

E-mail: sav@irgups.ru

**Abstract**

Current-quark masses are compared to the rest masses allowed by the Helmholtz equation in a polar model. Within the uncertainty of the current u quark mass determination, the current quark mass coincides with the rest mass allowed by the Helmholtz equation in the polar model in accordance with the second root of the zero Neumann function. Current d quark mass coincides with the rest mass calculated in accordance with the third root of the Bessel zero function.

On the basis of a comparison of these results with the results obtained earlier for ordinary real particles u and d quarks stability is discussed.

**Introduction**

Despite the limitations of relativistic quantum mechanics (RQM) [1-6] in comparison to quantum field theory (QFT) [2-4, 7, 8], the RQM equations in particular the Klein-Gordon and Dirac equations, continue to be actively investigated (e.g. [3-6, 9-14]).

The quark hypothesis was introduced in the QFT framework [15, 16] more than half a century ago. However, the quark confinement problem still remains unsolved [17, 18]. In this regard attempts to study quarks with the help of other possible approaches, in particular RQM methods, are of interest.

**Theoretical background**

It was suggested in [19] that a particle that does not interact with other particles can be described by the simplest wave equation

$$\left[\hat{p}^2 + \frac{\hbar^2}{c^2}\frac{\partial^2}{\partial t^2}\right]\Psi = 0, \qquad (1)$$

$\hat{p}^2 = (-i\hbar\nabla)^2$, $\nabla$ is a nabla operator, c is a light speed in a vacuum, $\hbar$ is the Planck constant. As a solution of this equation, we consider a wave function of the form

$$\Psi_{q,k} = \Psi_q \Psi_k. \qquad (2)$$

Wherein

$$\Psi_k = N_k \tilde{u} \exp\left[i\left(\mathbf{kr} - \frac{E}{\hbar}\right)t\right], \qquad (3)$$

$\Psi_q$ satisfies the Helmholtz equation

$$\hat{p}^2 \Psi_q = \hbar^2 q^2 \Psi_q. \qquad (4)$$

Then $\Psi_k$ satisfies the Klein-Gordon equation with the relativistic dispersion relation. In this case, the squares of the rest masses of particles with a relativistic dispersion are determined by the Helmholtz equation, that is, the square of the rest mass is an eigenvalue of the $\hat{p}^2/c^2$ operator in the rest frame.

Note that $\Psi_k$, $\Psi_{q,k}$ functions can be not only scalar but also multicomponent vector and spinor functions that transform according to irreducible representations of the Poincaré group, since equation (1) is invariant with respect to transformations of a 15-parameter conformal group containing as a subgroup the Poincare group [20]. In addition, it is well known that solutions of the Dirac equation are solutions of the Klein-Gordon equation, but not vice versa [21]. In turn, it can be seen from [19] that for a specific choice of a particular solution of the simplest wave equation, it falls into the Helmholtz equation and the Klein-Gordon equation. That is, it may be assumed that (1) describes particles with both integer and half-integer spins.

In spite of the fact that the state described by the wave function (2) is not a state with an eigenvalue of the momentum operator, for $\Psi_q, \Psi_k, \Psi_{q,k}$ functions when $q \perp k$, the relations

$$(\hat{p}\Psi_q)\hbar\kappa = 0, \quad (\hat{p}\Psi_q)(\hat{p}\Psi_k) = 0, \qquad (5)$$

$$\langle \Psi_{q,k}|\hat{p}|\Psi_{q,k}\rangle = \langle \Psi_k|\hat{p}|\Psi_k\rangle = \hat{p}\Psi_k = \hbar\mathbf{k} \qquad (6)$$

are valid. That is, in states described by wave functions $\Psi_k$ and $\Psi_{q,k}$ the average values of the momentum are the same. The Helmholtz equation (4) describes a particle in the rest frame for $\mathbf{k} = 0$, $\mathbf{v} = 0$, $\mathbf{v}$ is velocity of a particle, interpreted as the group velocity of the wave packet.

The polar model of the particle was considered in [19], in which (4) reduces to the Bessel equation. In this case, $\Psi_q$ contains a factor that is a linear combination of the Neumann and Bessel functions

$$F_{nl}(x) = B_{nl}^{(N)} N_n(x) + B_{nl}^{(B)} J_n(x). \qquad (7)$$

Here $J_n(x)$ is the Bessel function, $N_n(x)$ is the Neumann function, $B_{nl}^{(N)}$, $B_{nl}^{(B)}$ are expansion coefficients. The allowed discrete values of rest masses of particles are determined by the formula

$$m_{nl} = \pm \frac{1}{a}\frac{\hbar}{c} X_{nl}. \qquad (8)$$

Here $X_{nl}$ are the values of the roots of the $F_{nl}(x)$ functions in (7), $a$ is the constant entering the first boundary condition. Generally speaking, the $a$ value is arbitrary and it was chosen in such a way that (8) would give the electron mass at n = 0, $l$ = 1. Such a choice can be justified by the discovery of regularities in the mass spectrum calculated in accordance with the formula (8).

**Calculation results**

Here are the results of the calculation by the formula (8) with n = 0, $l$ = 2, 3, 4, 54.

$m_{0,2}^{(N)}$=2.2632 MeV/c², $m_{0,2}^{(B)}$=3.1567 MeV/c², $m_{0,3}^{(N)}$=4.0522 MeV/c², $m_{0,3}^{(B)}$=4.9487 MeV/c², $m_{0,4}^{(N)}$=5.8457 MeV/c², $m_{0,4}^{(B)}$=6.7431 MeV/c², $m_{0,54}^{(N)}$=95.666 MeV/c², $m_{0,54}^{(B)}$=96.565 MeV/c².

Upper indices (N) and (B) differ masses in the correspondence to the roots of the Neumann and Bessel functions.

Current-quark masses in a mass-independed substraction scheme [22] are as follows. For the u quark $m_u = 2.2^{+0.6}_{-0.4}$ MeV/c², for the d quark $m_d = 4.7^{+0.5}_{-0.4}$ MeV/c², for the s quark $m_s = 96^{+8}_{-4}$ MeV/c². As can be seen from a comparison of these quantities, within the limits of the determination of the mass of the "free" quarks [22], $m_{0,2}^{(N)}$ corresponds well to the $m_u$, $m_{0,3}^{(B)}$ corresponds well to the $m_d$, $m_{0,54}^{(N)}$ and $m_{0,54}^{(B)}$ are close in value to the $m_s$.

Because of the quasiperiodic character of the Bessel and Neumann functions in (7), the $|B_{nl}^{(B)}/B_{nl}^{(N)}|$ ratio is larger, the closer the value of the $F_{nl}(x)$ function to the value of the corresponding $X_{n\ell}^{(B)}$ root. And vice versa, $|B_{nl}^{(N)}/B_{nl}^{(B)}|$ is the more, the closer the value of the $F_{nl}(x)$ function (7) to the value of the corresponding $X_{n\ell}^{(N)}$ root. Since $m_{0,2}^{(N)}$ and $m_u$ coincide within the error of the $m_u$ definition, we can assume that $|B_{0,2}^{(B)}/B_{0,2}^{(N)}|$ in (7) for the u quark is practically zero, that is, the u-quark wave function includes the Neumann function and the Bessel function does not actually enter. For the d quark $m_{0,3}^{(B)} \approx m_d$ and vice versa, the wave function includes the Bessel function of zero order. The error in determining the mass of the "free" s quark [22] is too large (12 MeV/c²), and both calculated values are within the $m_s$ error. The errors in determining the mass of heavier quarks are even greater, so at present it makes no sense to compare their masses to the roots of the Bessel and Neumann functions.

**Discussions and conclusions**

It was noted in [19] that for the five real particles considered, the ratio of the expansion coefficients $B_{nl}^{(N)}$ and $B_{nl}^{(B)}$ in (7), in any case qualitatively, determines the lifetime of the real particles. The larger the contribution of the Neumann function that collapses at zero, the longer the lifetime. From this point of view, in the case of deconfinement, u quarks should be stable, and d quarks should have a short lifetime. This may be due, apparently, to different behaviors at zero ($x \to 0$) of the Bessel and Neumann functions. Since at present it is impossible to verify this assumption, the question arises of preserving the quark stability property in hadrons. Suppose that inside the hadron the u quark (N-like) provides more stability of the hadron than the d quark (B-like). This assumption is well satisfied if we compare the stability of a proton and a neutron. Indeed, the proton consisting of two u quarks and one d quark is more stable than the neutron consisting of one u quark and two d quarks. At the same time, this assumption is not satisfied for $\Delta$-baryons, which is apparently due to a smaller preservation of the quark individuality in these baryons in comparison with the proton and neutron.

For five real particles, as well as for the u quark and d quark in (7), it was sufficient to use only the zero (n = 0) Bessel and Neumann functions. Note that the zero Bessel function is the only one of all Bessel functions that does not vanish at $x \to 0$. This confirms the assumption that the stability of both ordinary real particles and quarks is due to the behavior of their single-particle wave functions at zero (for $x \to 0$).

The exact correspondence of the electron mass to the $X_{0,1}^{(N)}$ root of the Neumann function is due to the choice of the $a$ quantity. At the same time, the exact (within the error of determination) correspondence of the u quark mass to the $X_{0,2}^{(N)}$ root of the Neumann function and the correspondence of the mass of the d quark to the $X_{0,3}^{(B)}$ root of the Bessel function does not seem casual. Thus, when comparing the calculation results for the five ordinary real particles [19] and for the light quarks in the present paper, we see that the current-quark masses and the rest masses of real particles correspond to the roots of the Bessel and Neumann functions in the same way. Since the Helmholtz equation describes particles (and, apparently, quarks) in their rest frames ($\mathbf{k} = 0$), one can say that ordinary particles and quarks behave identically in their rest frames. This is another indirect evidence of the possibility of quarks deconfinement.